\documentclass[12pt]{article}
\begin{document}
\begin{center}
{\large\bf Lorentz Violation of Quantum Gravity} \vskip 0.3 true in {\large J.
W. Moffat} \vskip 0.3 true in {\it The Perimeter Institute for
Theoretical Physics, Waterloo, Ontario, N2J 2W9, Canada} \vskip
0.3 true in and \vskip 0.3 true in {\it Department of Physics,
University of Waterloo, Waterloo, Ontario N2Y 2L5, Canada}
\end{center}
\begin{abstract}%
A quantum gravity theory which becomes renormalizable at short distances due to a spontaneous symmetry breaking of Lorentz invariance and diffeomorphism invariance is studied. A breaking of Lorentz invariance with the breaking patterns $SO(3,1)\rightarrow O(3)$ and $SO(3,1)\rightarrow O(2)$, describing $3+1$ and $2+1$ quantum gravity, respectively, is proposed. A complex time dependent Schr\"odinger equation (generalized Wheeler-DeWitt equation) for the wave function of the universe exists in the spontaneously broken symmetry phase at Planck energy and in the early universe, uniting quantum mechanics and general relativity. An explanation of the second law of thermodynamics and the spontaneous creation of matter in the early universe can be obtained in the symmetry broken phase of gravity.
\end{abstract}
\vskip 0.2 true in
e-mail: john.moffat@utoronto.ca


\section{Introduction}

The quantum theory of gravitation in four dimensions ($D=4$) quantized on a fixed background such as Minkowski spacetime with the metric $\eta_{\mu\nu}={\rm diag}(+1,-1,-1,-1)$ is not renormalizable~\cite{tHooft,Goroff}. This has led to much effort to search for a physically consistent and finite quantum gravity theory. Many attempts include string theory~\cite{Polchinski}, loop quantum gravity~\cite{Ashtekar,Rovelli,Smolin}, and a finite non-local, regularized quantum gravity theory~\cite{Moffat}. In earlier papers, the local Lorentz and diffeomorphism invariance of gravity was spontaneously broken in a {\it vierbein} gauge theory~\cite{Moffat2,Moffat3}. From the symmetry breaking schemes $SO(3,1)\rightarrow O(3)$ and $SO(3,1)\rightarrow O(2)$, quantum gravity theory underwent a reduction to a $3+1$ and $2+1$ theory, respectively. It was assumed that the symmetry broken phase occurred at an energy $E \sim E_P$ where $E_P=1/M_P\sim 10^{19}$ GeV is the Planck energy. This would correspond to a breaking of the symmetry at a critical temperature $T\sim T_c\sim E_P$, in the very early universe. The reduction of the quantum gravity to lower-dimensional gravity theories at very short distances can lead to a renormalizable quantum gravity theory~\cite{Visser,Townsend}.

Recently, the idea of reducing quantum gravity to a $3+1$ theory has been revived by Horava~\cite{Horava}, who based the Lorentz violation on an ``anisotropic scaling" of the space and time dimensions. The idea is to introduce a quantity $Z$, with the physical dimensions: $[Z]=[dx]^z/[dt]$ and for the relativistic gravity theory $Z\rightarrow 1$. We argue that the Lorentz and diffeomorphism violation of gravity has a more intuitive and physical basis in spontaneous symmetry breaking of the gravitational action. The spontaneous symmetry breaking mechanism only breaks the vacuum or ground state of the gravitational system, retaining a "hidden" local gauge invariance symmetry of the action that preserves Takahashi-Ward identities and other attractive properties of the purely gauge invariant formalism.

In the symmetry broken phase and in the $3+1$ or $2+1$-dimensional quantum gravity, time becomes ``absolute" and is described by the $R\times O(3)$ Lema\^{i}tre-Friedman-Robertson-Walker (LFRW) cosmology. The breaking of time translational and Lorentz invariance leads to a complex Schr\"odinger equation (generalized Wheeler-DeWitt equation~\cite{DeWitt}) for the wave function of the universe, thereby solving the problem of time and uniting quantum mechanics and relativistic gravity~\cite{Moffat2}.

The spontaneous symmetry breaking mechanism in the {\it vierbein} gauge formalism has 3 massless degrees of freedom associated with the $O(3)$ rotational invariance, and 3 massive degrees of freedom associated with the broken Lorentz ``boosts". The massive quantum gravity in $3+1$ dimensions can satisfy unitarity and be renormalizable, in contrast to the $D=4$ quantum gravity which will violate unitarity if renormalizable~\cite{Stelle}. Moreover, the $2+1$ quantum gravity in which both local Lorentz invariance and rotational invariance are broken can for a massive graviton be unitary, ghost-free and renormalizable~\cite{Townsend}.

The spontaneous symmetry broken phase will induce a violation of conservation of energy and explain the generation of matter in the very early universe. Moreover, in the ordered symmetry broken phase entropy will be at a minimum. After the phase transition as the Universe expands into the disordered phase with $SO(3)\rightarrow SO(3,1)$ or $O(2)\rightarrow SO(3,1)$ there will be a large increase in entropy with an arrow of time created by the spontaneous choice of symmetry breaking.

\section{Spontaneous Symmetry Breaking of Gravity}

Let us define the metric in any non-inertial coordinate system by
\begin{equation}
\label{vierbeinmetric}
g_{\mu\nu}(x)=e^a_{\mu}(x)e^b_{\nu}(x)\eta_{ab},
\end{equation}
where
\begin{equation}
e^a_{\mu}(X)=\biggl({\partial \zeta^a_X(x)\over \partial
x^{\mu}}\biggr)_{x=X}.
\end{equation}

The $\zeta^a_X$ are a set of locally inertial coordinates at $X$. The
vierbeins $e^a_{\mu}$ satisfy the orthogonality relations:
\begin{equation}
e^a_{\mu}e_b^{\mu}=\delta^a_b,\quad e^{\mu}_ae_{\nu}^a=\delta^{\mu}_{\nu},
\end{equation}
which allow us to pass from the flat tangent space coordinates (the fibre
bundle tangent space) labeled by $a,b,c...
$ to the the world spacetime coordinates (manifold) labeled by $\mu,\nu,
\rho...$.
The fundamental form (\ref{vierbeinmetric}) is invariant under Lorentz transformations:
\begin{equation}
e^{\prime\,a}_{\mu}(x)=L^a_b(x)e^b_{\mu}(x),
\end{equation}
where $L^a_b(x)$ are the homogeneous $SO(3,1)$ Lorentz transformation
coefficients that can depend on position in spacetime, and which satisfy
\begin{equation}
L_{ac}(x)L^a_d(x)=\eta_{cd}.
\end{equation}

For a general field $f_n(x)$ the transformation rule will take the form
\begin{equation}
\label{Dtransformation}
f_n(x)\rightarrow \sum_m[D(L)(x)]_{nm}f_m(x),
\end{equation}
where $D(L)$ is a matrix representation of the (infinitesimal) Lorentz group.

The $e^a_{\mu}$ will satisfy
\begin{equation}
e^a_{\mu,\sigma}+(\Omega_\sigma)^a_ce^c_{\mu}-\Gamma^{\rho}_{\sigma\mu}
e^a_{\rho}=0,
\end{equation}
where $e^a_{\mu ,\nu}=\partial e^a_{\mu}/\partial x^{\nu}$, $\Omega_{\mu}$ is
the spin connection of gravity and
$\Gamma^{\lambda}_{\mu\nu}$ is the Christoffel connection. Solving for
$\Gamma$ gives
\begin{equation}
\Gamma_{\sigma\lambda\rho}=g_{\delta\rho}\Gamma^{\delta}_{\sigma\lambda}
=\eta_{ab}(D_{\sigma}e^a_{\lambda})e^b_{\rho},
\end{equation}
where
\begin{equation}
D_{\sigma}e^a_{\mu}=e_{\mu,\sigma}^a+(\Omega_{\sigma})^a_ce^c_{\mu}
\end{equation}
is the covariant derivative operator with respect to the gauge connection
$\Omega_{\mu}$. By differentiating
(\ref{vierbeinmetric}), we get
\begin{equation}
g_{\mu\nu,\sigma}-g_{\rho\nu}\Gamma^{\rho}_{\mu\sigma}-g_{\mu\rho}
\Gamma^{\rho}_{\nu\sigma}=0,
\end{equation}
where we have used $(\Omega_{\sigma})_{ca}=-(\Omega_{\sigma})_{ac}$.

The (spin) gauge connection $\Omega_{\mu}$ remains invariant under the Lorentz
transformations provided:
\begin{equation}
\label{Spintransformation}
(\Omega_{\sigma})^a_b\rightarrow [L\Omega_{\sigma}L^{-1}
-(\partial_{\sigma}L)L^{-1}]^a_b.
\end{equation}
A curvature tensor can be defined by
\begin{equation}
([D_{\mu},D_{\nu}])^a_b=(R_{\mu\nu})^a_b,
\end{equation}
where
\begin{equation}
(R_{\mu\nu})^a_b=(\Omega_{\nu})^a_{b,\mu}-(\Omega_{\mu})^a_{b,\nu}
+([\Omega_{\mu},\Omega_{\nu}])^a_b.
\end{equation}

The curvature tensor transforms like a gauge field strength:
\begin{equation}
(R_{\mu\nu})^a_b\rightarrow L^a_c(R_{\mu\nu})^c_d(L^{-1})^d_b.
\end{equation}

In holonomic coordinates, the curvature tensor is
\begin{equation}
R^{\lambda}_{\sigma\mu\nu}=(R_{\mu\nu})^a_be^{\lambda}_ae^b_{\sigma}
\end{equation}
and the scalar curvature takes the form
\begin{equation}
R=e^{\mu a}e^{\nu b}(R_{\mu\nu})_{ab}.
\end{equation}

At the Planck energy $E_P$ the local Lorentz vacuum symmetry is spontaneously broken. We postulate
the existence of a field, $\phi$, and assume that the
vacuum expectation value (vev) of the field, $<\phi>_0$, will vanish for
for $E=E_c < E_P$ or at a temperature $T < T_c\sim M_P$, when
the local Lorentz symmetry is restored. At $E\sim E_P$ the non-zero
vev will break the symmetry of the ground state of the Universe from $SO(3,1)$
down to $O(3)$ or $O(2)$. The domain formed by the direction of the vev of the
$\phi$ field will produce a time arrow pointing in the direction of increasing
entropy and the expansion of the Universe.

Let us introduce the fields $\phi^a(x)$ which are invariant
under Lorentz transformations
\begin{equation}
\phi^{\prime a}(x)=L^a_b(x)\phi^b(x).
\end{equation}
We can use the {\it vierbein} to convert $\phi^a$ into a 4-vector in coordinate
space: $\phi^{\mu}=e^{\mu}_a\phi^a$.
The covariant derivative operator acting on $\phi$ is defined by
\begin{equation}
D_{\mu}\phi^a=[\partial_{\mu}\delta^a_b+(\Omega_{\mu})^a_b]\phi^b.
\end{equation}

If we consider infinitesimal Lorentz transformations
\begin{equation}
L^a_b(x)=\delta^a_b+\omega^a_b(x)
\end{equation}
with
\begin{equation}
\omega_{ab}(x)=-\omega_{ba}(x),
\end{equation}
then the matrix D in (\ref{Dtransformation}) has the form:
\begin{equation}
D(1+\omega (x))=1+{1\over 2}\omega^{ab}(x)\sigma_{ab},
\end{equation}
where the $\sigma_{ab}$ are the six generators of the Lorentz group
which satisfy $\sigma_{ab}=-\sigma_{ba}$ and the commutation relations
\begin{equation}
[\sigma_{ab},\sigma_{cd}]=\eta_{cb}\sigma_{ad}-\eta_{ca}\sigma_{bd}
+\eta_{db}\sigma_{ca}-\eta_{da}\sigma_{cb}.
\end{equation}

The set of fields $\phi$ transforms as
\begin{equation}
\phi^{\prime}(x)=\phi(x)+\omega^{ab}(x)\sigma_{ab}\phi(x).
\end{equation}
The gauge spin connection which satisfies the transformation law (\ref{Spintransformation}) is
given by
\begin{equation}
\Omega_{\mu}={1\over 2}\sigma^{ab}e^{\nu}_ae_{b\nu;\mu},
\end{equation}
where ; denotes covariant differentiation with respect to the Christoffel connection. We
introduce a spontaneous symmetry breaking
sector into the Lagrangian density such that the gravitational vacuum symmetry, which we
set equal to the Lagrangian symmetry at low temperatures, will break to a smaller symmetry at high
temperature. The breaking of the symmetry at a higher temperature is an example of ``anti-restoration'' symmetry
breaking~\cite{Moffat2}. The vacuum symmetry breaking leads to the interesting possibility that exact zero
temperature conservation laws e.g. electric charge and baryon number are broken in the early
Universe. In our case, we shall find that the spontaneous breaking of the
Lorentz symmetry of the vacuum leads to a violation of the exact zero
temperature conservation of energy in the early Universe~\cite{Moffat2}.

Consider the potential:
\begin{equation}
V(\phi)=\biggl[\lambda\sum_{a=0}^3\phi^a\phi_a-{1\over 2}\mu^2\biggr]
\sum_{b=0}^3\phi^b\phi_b,
\end{equation}
where $\lambda > 0$ is a coupling constant such that $V(\phi)$ is
bounded from below. Our Lagrangian density takes the form~\cite{Moffat2,Moffat4,Kostelecky}:
\begin{equation}
\label{Lagrangian}
{\cal L}={\cal L}_G -\sqrt{-g}\biggl[{1\over 4}B^{ab}B_{ab}
+V(\phi)\biggr],
\end{equation}
where
\begin{equation}
B_{ab}=D_b\phi_a-D_a\phi_b,
\end{equation}
and
\begin{equation}
{\cal L}_G=-\frac{1}{16\pi G}\int d^4xe[R(\Omega)-2\Lambda].
\end{equation}
Moreover, $e\equiv\sqrt{-g}={\rm det}(e^a_\mu e_{a\nu})^{1/2}$, $R(\Omega)$ denotes the scalar curvature determined by
the spin connection and $\Lambda$ is the cosmological constant.

If $V$ has a minimum at $\phi_a=v_a$, then the spontaneously broken solution
is given by $v_a^2=\mu^2/4\lambda$ and an expansion of $V$ around the
minimum yields the mass matrix:
\begin{equation}
\label{massmatrix}
(\mu^2)_{ab}={1\over 2}\biggl({\partial^2 V\over \partial \phi_a \partial
\phi_b}\biggr)_{\phi_a=v_a}.
\end{equation}
We can choose $\phi_a$ to be of the form
\begin{equation}
\phi_a=\left(\matrix{0\cr
0\cr
0\cr
v\cr}\right)=\delta_{a0}
(\mu^2/4\lambda)^{1/2}.
\end{equation}
All the other solutions of $\phi_a$ are related to this one by a Lorentz
transformation. Then, the homogeneous Lorentz group $SO(3,1)$ is broken
down to the
spatial rotation group $O(3)$. The three rotation generators $J_i
(i=1,2,3)$ leave the vacuum invariant
\begin{equation}
J_iv_i=0,
\end{equation}
while the three Lorentz-boost generators $K_i$ break the vacuum symmetry
\begin{equation}
K_iv_i\not= 0.
\end{equation}
The $J_i$ and $K_i$ satisfy the commutation relations
\begin{equation}
[J_i,J_j]=i\epsilon_{ijk}J_k,\quad [J_i,K_j]=i\epsilon_{ijk}K_k,\quad
[K_i,K_j]=-i\epsilon_{ijk}K_k.
\end{equation}

The mass matrix $(\mu^2)_{ab}$ can be calculated from (\ref{massmatrix}):
\begin{equation}
(\mu^2)_{ab}=(-{1\over 2}\mu^2+2\lambda v^2)\delta_{ab}+4\lambda v_av_b
=\mu^2\delta_{a0}\delta_{b0},
\end{equation}
where $v$ denotes the magnitude of $v_a$. There are three zero-mass Nambu-Goldstone
bosons, the same as the number of massive bosons, and there are three
massless degrees of freedom corresponding to the unbroken $O(3)$ symmetry.  After
the spontaneous breaking of the vacuum, one massive physical particle
$\Phi$ remains. No ghost particles will occur in the unitary gauge.
The mass term in the Lagrangian density is given in the unitary gauge by
\begin{equation}
{\cal L}_M={1\over 2}\sqrt{-g}v_bv_c(\Omega_{\mu})^{ab}(\Omega^{\mu})^c_a
={1\over 2}\sqrt{-g}(\mu^2/4\lambda)\sum_{i=1}^3((\Omega_{\mu})^{i0})^2.
\end{equation}
When Lorentz symmetry is restored for $E < E_c$, then $v=0$ and ${\cal L}_M=0$ and
we obtain the standard GR Lagrangian density with a massless spin-2 graviton, coupled minimally to
a spin-1 particle.

We could have extended this symmetry breaking pattern to the case where we
have two sets of vector field representations, $\phi_{a1}$ and $\phi_{a2}$. The
invariant spin connection can depend on the length of each vector
and the angle between them, $\vert\phi_{a1}\vert, \vert\phi_{a2}\vert$, and
$\chi=\vert \phi_{a1}\phi^a_2\vert$. The solutions for the minimum must be
obtained from the conditions imposed on these three quantities. We can choose
$\phi_{a1}$ with only the last component non-zero and $\phi_{a2}$ with the
last two components non-zero in order to satisfy these conditions. The
Lorentz $SO(3,1)$ symmetry is then broken down to $O(2)$ (or $U(1)$)
symmetry~\cite{Li}.

A phase transition is assumed to occur at the critical temperature $T_c$,
when $v_a\not= 0$ and the Lorentz symmetry is broken and the three gauge
fields $(\Omega_{\mu})^{i0}$ become massive degrees of freedom. Below
$T_c$ the Lorentz symmetry is restored, and we
regain the usual classical gravitational field with massless gauge fields
$\Omega_{\mu}$. The symmetry breaking will extend to the
singularity or the possible singularity-free initial state at $t=0$,
and since quantum effects associated with gravity do not become important
before $E_P$, we expect that $E_c\sim 10^{19}$ GeV.

A calculation of the effective potential for the symmetry breaking contribution in
(\ref{Lagrangian}) shows that extra minima in the potential $V(\phi)$ can occur for a
noncompact group such as $SO(3,1)$. This fact has been explicitly demonstrated
in a model with $O(n)\times O(n)$ symmetric four-dimensional $\phi^4$ field
theory~\cite{Salmonson}. This model has two irreducible representations of fields, ${\vec
\phi}_1$ and ${\vec \phi}_2$, transforming as (n,1) and (1,n), respectively.
The potential is
\begin{equation}
V=\sum_i{1\over 2}m_i^2{\vec\phi}^2_i+\sum_{i,j}{1\over 8}{\vec\phi}_i^2
\lambda_{ij}{\vec\phi}_j^2.
\end{equation}
The requirement of boundedness from below gives $(\lambda_{12}=\lambda_{21})$:
\begin{equation}
\lambda_{11} > 0,\quad \lambda_{22} >-(\lambda_{11}\lambda_{22})^{1/2}.
\end{equation}

If we have $\lambda_{12} < -(1+2/n)\lambda_{22}$, then the one-loop free energy
predicts spontaneous symmetry breaking to $O(n)\times O(n-1)$ at
sufficiently high temperatures without symmetry breaking at small temperatures.
The standard symmetry breaking restoration theorems can be broken in this
case because the dynamical variables ${\vec \phi}_i$ do not form a compact
space$^{}$.

After the symmetry is restored for $E < E_P$, the entropy will rapidly increase provided that no further
phase transition occurs which breaks the Lorentz symmetry of the vacuum.
Thus, the symmetry breaking mechanism explains in a natural way the low entropy at the initial
state at $t\sim 0$  and the large entropy in the present universe.

Since the ordered phase is at a much lower entropy than the disordered phase and due
to the existence of a domain determined by the direction of the vev of the
$\phi$ field, a natural explanation is given for the cosmological arrow of
time and the origin of the second law of thermodynamics.
Thus, the spontaneous symmetry breaking of the gravitational
vacuum corresponding to the breaking pattern, $SO(3,1)\rightarrow O(3)$,
leads to a manifold with the structure $R\times O(3)$, in which time
appears as an absolute external parameter~\cite{Moffat2}. The vev, $<\phi>_0$, points in a chosen direction
of time to break the symmetry creating an arrow of time. The evolution from a state of low
entropy in the ordered phase to a state of high entropy in the disordered phase explains
the second law of thermodynamics.

\section{$3+1$ Quantum Gravity}

The action in Einstein's gravitational theory
for a fixed three-geometry on a boundary is~\cite{Arnowitt,DeWitt}:
\begin{equation}
S_E={1\over 16\pi G}\biggl[\int_{\partial M}d^3xh^{1/2}2K+\int_Md^4x(-g)^{1/2}
(R+2\Lambda)\biggr],
\end{equation}
where the second term is integrated over spacetime and the first over its
boundary, $K$ is the trace of the extrinsic curvature $K_{ij}$ (i,j=1,2,3)
of the boundary three-surface.
We write the metric in the usual $3+1$ form:
\begin{equation}
ds^2=(N^2-N_iN^i)dt^2-2N_idx^idt-h_{ij}dx^idx^j,
\end{equation}
and the action becomes
\begin{equation}
S_E={1\over 16\pi G}\int d^4xh^{1/2}N[-K_{ij}K^{ij}+K^2-R(h)^{(3)}+2\Lambda],
\end{equation}
where
\begin{equation}
K_{ij}={1\over N}\biggl[-{1\over 2}{\partial h_{ij}\over \partial t}
+N_{(i\vert j)}\biggr].
\end{equation}
$R^{(3)}$ denotes the scalar curvature constructed from the three-metric
$h_{ij}$ and a stroke denotes the covariant derivative with respect to the
latter quantity. The matter action $S_M$ can also be constructed from
the $N, N_i, h_{ij}$ and the matter field.

The super-Hamiltonian density is given by
\begin{equation}
H=NH_0+N^iH_i=H_0\sqrt h +N^iH_i,
\end{equation}
where $H_0$ and $H_i$ are the usual Hamiltonian and momentum constraint
functions, defined in terms of the canonically conjugate momenta $\pi ^{ij}$
to the dynamical variables $h_{ij}$:
\begin{equation}
\pi^{ij}={\delta L_E\over \delta (\partial {h}_{ij}/\partial t)},
\end{equation}
where $L_E$ is the Einstein-Hilbert Lagrangian density. Classically, the Dirac
constraints are
\begin{equation}
\label{constraints}
H_i=0,\quad H_0=0.
\end{equation}
These constraints are a direct consequence of the general covariance of
Einstein's theory of gravity.

In quantum mechanics, a suitably normalized wave function is defined by the
path integral
\begin{equation}
\label{pathintegral}
\psi ({\vec x},t)=-\int [d{\vec x}(t)]\hbox{exp}[iS({\vec x}(t))].
\end{equation}
We obtain
\begin{equation}
{\partial \psi\over \partial t}=-i\int [d{\vec x}(t)]
{\partial S\over \partial t}\hbox{exp}(iS),
\end{equation}
which leads to the Schr\"odinger equation
\begin{equation}
i{\partial \psi\over \partial t}=H\psi.
\end{equation}

We define the wave function of the universe to be~\cite{Hawking,Moffat2}:
\begin{equation}
\label{wavefunction}
\Psi[h_{ij},\phi]=-\int [dg][d\phi]\mu[g,\phi]\exp(iS[g,\phi]),
\end{equation}
where $\phi$ denotes a matter field, $S$ is the total action and $\mu[g,\phi]$ is an invariant
measure. The integral or sum is over a class of spacetimes with a compact boundary on which
the induced metric $h_{ij}$ and field configurations match $\phi$ on the boundary.
A differential equation for the wave function of the Universe, $\Psi$, can be
derived by varying the end conditions
on the path integral (\ref{wavefunction}). Since the theory is diffeomorphism invariant the
wave function is independent of time and we obtain
\begin{equation}
{\delta \Psi\over \delta N} = -i\int [dg][d\phi]\mu[g,\phi]
\biggl[{\delta S\over \delta N}\biggr]\hbox{exp}(iS[g,\phi])=0,
\end{equation}
where we have taken into account the translational invariance of the measure
factor $\mu[g,\phi]$. Thus, the value of the integral is left
unchanged by an infinitesimal translation of the integration variable $N$
and leads to the operator equation:
\begin{equation}
H_0\Psi=0.
\end{equation}

The classical Hamiltonian constraint equation takes the form
\begin{equation}
H_0=\delta S/\delta N = h^{1/2}(-K^2+K_{ij}K^{ij}-R^{(3)}+2\Lambda
+16\pi GT_{nn})=0,
\end{equation}
where $T_{nn}$ is the stress-energy tensor of the matter field projected
in the direction normal to the surface. By a suitable factor ordering (ignoring
the well-known ``factor ordering" problem), the
classical equation $\delta S/\delta N = 0$ translates into the operator
identity
\begin{equation}
\biggl\{-\gamma_{ijkl}{\delta^2\over \delta h_{ij}\delta h_{kl}}+h^{1/2}
\biggl[R^{(3)}(h)-2\Lambda -{16\pi\over M_P^2}
T_{nn}\biggl(-i{\delta\over \delta\phi},
\phi\biggr)\biggr]\biggr\}\Psi[h_{ij},\phi]=0,
\end{equation}
where $\gamma_{ijkl}$ is the metric on superspace,
\begin{equation}
\gamma_{ijkl}={1\over 2}h^{-1/2}(h_{ik}h_{jl}+h_{il}h_{jk}-h_{ij}h_{kl}).
\end{equation}
This is the familiar Wheeler-DeWitt equation for a closed universe~\cite{DeWitt}.

We would expect that the wave
function of the universe should be time dependent and lead to a complex
Schr\"odinger equation or its covariant counterpart-- the Tomonaga-Schwinger
equation:
\begin{equation}
i{\delta \Psi\over \delta \tau} = {\cal H}\Psi,
\end{equation}
which leads to the ordinary time dependent Schr\"odinger wave equation for
global time variations,
with a positive-definite probabilistic interpretation. We
therefore propose a new definition of the wave function
of the universe which takes the form~\cite{Moffat2}:
\begin{equation}
\Psi[h_{ij},\phi]=-\int [dg][d\phi]M[g,\phi]\hbox{exp}(iS[g,\phi]),
\end{equation}
where $M[g,\phi]$ is a measure factor that breaks the time
translational invariance of the path integral and makes the wave function
$\Psi$ explicitly time dependent. We now obtain
\begin{equation}
{\delta \Psi\over \delta N} =-\int [dg][d\phi]{\delta M\over \delta N}
\hbox{exp}(iS)-i\int [dg][d\phi]M[g,\phi]{\delta S
\over \delta N}\hbox{exp}(iS).
\end{equation}
This leads to the time dependent Schr\"odinger equation
\begin{equation}
i{\delta \Psi\over \delta N}=\tilde {H}_0\Psi,
\end{equation}
where $\tilde {H}_0$ denotes
\begin{equation}
\tilde {H}_0=
- i{\delta \hbox{ln}M\over \delta N}.
\end{equation}

A simple example of a measure factor that brings in an explicit time
dependence (or $N$ dependence) is
\begin{equation}
M[g,\phi ]=\mu [g,\phi]N^b.
\end{equation}
This measure factor $M[g,\phi]$ retains the momentum
constraint equation $H_i=0$ as an operator equation:
\begin{equation}
H_i\Psi=0,
\end{equation}
while keeping the invariance
of the spatial three-geometry at the quantum mechanical level as well as
at the classical level. If the measure $M[g,\phi]$ is chosen so that the
diffeomorphism group ${\cal D}$ is broken down to a sub-group ${\cal S}$, then
there will exist a
minimal choice of $M[g,\phi]$ which will break time translational invariance.
The choice of $M[g,\phi]$ is not unique and some, as yet, unknown physical
principle is needed to determine $M[g,\phi]$.
At the classical level, we continue to maintain general covariance
and the classical constraint equations (\ref{constraints}) hold. The Bianchi identities
\begin{equation}
{{G_{\mu}}^{\nu}}_{;\nu}=0,
\end{equation}
are valid, where ${G_{\mu}}^{\nu}={R_{\mu}}^{\nu}-
{1\over 2}{\delta_{\mu}}^{\nu}R$. It is only the quantum
mechanical wave function that breaks the diffeomorphism invariance i.e.,
$N$ is no longer a free variable for the wave function of the universe.
This leads naturally to a cosmic time which can be used to measure time
dependent quantum mechanical observables. We find that for any operator
$O$, we get
\begin{equation}
\label{operatoreq}
{\delta\over \delta N} <O> =i<[H,O]>,
\end{equation}
which constitutes the quantum mechanical version of Hamilton's equation.
In contrast to the Wheeler-DeWitt equation, Ehrenfest's theorem follows directly from
(\ref{operatoreq}).

\section{Conclusions}

We have succeeded in arriving at a unification of quantum mechanics and gravity within a conceptually
logical picture, since both of these pillars of modern physics are with us to
stay. However, to achieve this we have postulated that Poincar\'e invariance and diffeomorphism invariance are
violated at the Planck energy $E_P$. There exist stringent experimental bounds on violation of Lorentz invariance at lower
energies~\cite{Taylor}, but there is no observational evidence that Lorentz invariance is strictly maintained at
the Planck energy.

In $3+1$ and $2+1$ gravity the power counting of momenta in Feynman loop graphs allows the quantum gravity to be
renormalizable~\cite{Visser}. Moreover, the problem of time in general relativity that prevents
a logically consistent solution to uniting quantum mechanics and gravity is also resolved. This would lead
one to believe that spontaneously breaking Lorentz symmetry at the Planck energy $E_P$ could be a satisfactory
solution to quantum gravity. To confirm that this way to resolve the problem of quantum gravity is realized in nature,
it is necessary to experimentally detect a violation of Poincar\'e invariance at the Planck energy.

\section{Acknowledgments}

I thank Viktor Toth for stimulating and helpful discussions. This work was supported by the Natural Science and Engineering
Research Council of Canada. Research at the Perimeter Institute for Theoretical Physics is supported by the Government of
Canada through NSERC and the Province of Ontario through the Ministry of Research and Innovation (MRI).


\begin{thebibliography}{1}

\bibitem{tHooft} G. 't Hooft and M. Veltman, Annales de l'institut Henri Poincaré (A) Physique Th\'eorique,
{\bf 20} 69 (1974).

\bibitem{Goroff} M. H. Goroff and A. Sagnotti, Phys. Lett. {\bf 160B}, 81 (1985).

\bibitem{Polchinski} J. Polchinski, {\it String Theory}, Cambridge University Press, 1998.

\bibitem{Ashtekar} A. Ashtekar, arXiv:0705.2222 (2007).

\bibitem{Rovelli} C. Rovelli, {\it Quantum Gravity}, Cambridge University Press, 2004.

\bibitem{Smolin} L. Smolin, {\it Three Roads to Quantum Gravity}, Basic Books, 2002.

\bibitem{Moffat} J. W. Moffat, arXiv:hep-ph/0102088.

\bibitem{Moffat2} J. W. Moffat, Found.Phys. {\bf 23}, 411 (1993);
arXiv:gr-qc/9209001;

\bibitem{Moffat3} J. W. Moffat, Int. J. Mod.Phys. {\bf D2}, 351 (1993), arXiv:gr-qc/9211020.

\bibitem{Visser}  M. Visser, arXiv:hep-th/0902.0590.

\bibitem{Townsend}  E. A. Bergshoeff, O. Hohm and P. K. Townsend, arXiv:hep-th/0901.1766; arXiv:hep-th/0905.1259.

\bibitem{Horava} P. Horava, Phys. Rev. {\bf D79}, 084008 (2009), arXiv: hep-th/0902.3657; P. Horava, arXiv:hep-th/0902.3657.

\bibitem{DeWitt} B. S. DeWitt, Phys. Rev. {\bf 160}, 1113 (1967); J. A. Wheeler, {\it Battelle
Rencontres}, eds. C. DeWitt and J. A. Wheeler, Benjamin, New York, 1968.

\bibitem{Stelle} K. S. Stelle, Phys. Rev. {\bf D16}, 953 (1977).

\bibitem{Moffat4} J. W. Moffat, Int. J. Mod.Phys. {\bf D12}, 1279 (2003), arXiv:hep-th/0211167.

\bibitem{Kostelecky} V. A. Kostelecky and S. Samuel, Phys. Rev. {\bf D40}, 1886 (1989).

\bibitem{Arnowitt} R. L. Arnowitt, S. Deser and C. W. Misner, {\it Gravitation: An Introduction to Current Research}
ed. L. Witten, Wilew, 1962, arXiv:gr-qc/0405109.

\bibitem{Hawking} J. B. Hartle and S. W. Hawking, Phys. Rev. {\bf D28}, 2960 (1983).

\bibitem{Li} Ling-Fong Li, Phys. Rev. {\bf D9}, 1723 (1974).

\bibitem{Salmonson} P. Salmonson and B.K. Skagerstam, Phys. Letts. {\bf B155}, 98 (1985).

\bibitem{Taylor} L. Maccione, A. M. Taylor, D. Mattingly and S. Liberati, arXiv:astro-ph/0902.1756.


\end{thebibliography}

\end{document}